# A new approach to the thermodynamic analysis of gas power cycles

Di HE[1], Zhipeng DUAN[1], Chaojun WANG[1], Boshu HE[1,2] *

*[1]Institute of Combustion and Thermal Systems, School of Mechanical, Electronic and Control Engineering, Beijing Jiaotong University, Beijing 100044, China

[2]School of Mechanical and Power Engineering, Cangzhou Jiaotong College, Huanghua 061199, Hebei Province, China

* Corresponding author; E-mail: hebs@bjtu.edu.cn

Abstract: Engineering Thermodynamics has been the core course of many science and engineering majors around the world, including energy and power, mechanical engineering, civil engineering, aerospace, cryogenic refrigeration, food engineering, chemical engineering, and environmental engineering, among which gas power cycle is one of the important contents. However, many Engineering Thermodynamics textbooks focus only on evaluating the thermal efficiency of gas power cycle, while the important concept of specific cycle work is ignored. Based on the generalized temperature-entropy diagram for the gas power cycles proposed by the authors, an ideal Otto cycle and an ideal Miller-Diesel cycle are taking as examples for the thermodynamic analyses of gas power cycles. The optimum compression ratio (or the pressure ratio) for the maximum specific cycle work or the maximum mean effective pressure is analyzed and determined. The ideal Otto and the ideal Miller-Diesel cycles, and also other gas power cycles for movable applications, are concluded that the operation under the maximum specific cycle work or the maximum mean effective pressure, instead of under the higher efficiency, is more economic and more reasonable. We concluded that the very important concept, i.e., the optimum compression (or pressure) ratio for the gas power cycles, should be emphasized in the Engineering Thermodynamics teaching process and in the latter revised or the newly edited textbooks, in order to better guide the engineering applications.

## 1. Introduction

Çengel et al. [1-4] wrote in the Preface of the textbooks: Thermodynamics is an exciting and fascinating subject that deals with energy, and thermodynamics has long been an essential part of engineering curricula all over the world. It has a broad application area ranging from microscopic organisms to common household appliances, transportation vehicles, power generation systems, and even philosophy. This implies this curriculum is essential for the engineering discipline.

Moran et al. [5-7] wrote in their textbooks: Engineers use principles drawn from thermodynamics and other engineering sciences, including fluid mechanics and heat and mass transfer, to analyze and design devices intended to meet human needs. Throughout the twentieth century, engineering applications of thermodynamics helped pave the way for significant improvements in our quality of life with advances in major areas such as surface transportation, air travel, space flight, electricity generation and transmission, building heating and cooling, and improved medical practices. In the twenty-first century, engineers will create the technology needed to achieve a sustainable future. Thermodynamics will continue to advance human well-being by addressing looming societal challenges owing to declining supplies of energy resources: oil, natural gas, coal, and fissionable material; effects of global climate change; and burgeoning population. This also means this curriculum is essential for the engineering discipline.

That is to say that Engineering Thermodynamics is quite important both for science and engineering. It is a science that studies the principles of the conversion and the application among thermal energy and mechanical energy. One of the main tasks of Engineering Thermodynamics is to clarify theoretically the ways to improve the efficiency of heat engine and the utilization rate of heat energy.

Along with the progress of science and technology and production development, the scope of the research and application of Engineering Thermodynamics is not limited to just as a heat engine (or refrigeration) theoretical basis. It has now expanded to many fields of engineering technology, such as aerospace, high-energy laser, heat pump, air separation, air conditioning, water desalination, chemical refining, biological engineering, low temperature superconducting, physical chemistry, etc. All the fields need the basic theory of Engineering Thermodynamics and the guidance of the basic knowledge. Therefore, Engineering Thermodynamics has become a compulsory basic course for many related majors.

From the perspective of guiding engineering application, Engineering Thermodynamics plays an important theoretical guiding role in guiding the application related to thermal energy and a very important role in the design and transformation of equipment, and has the significance of guiding light for the efficient use of energy. It is no exaggeration to say that without the correct theoretical system of Engineering Thermodynamics, there would be no modern advanced power equipment.

In this regard, Engineering Thermodynamics textbooks have played a very important role. Currently, typical textbooks in English used by high-level universities include [1-13], represented by textbooks edited by Çengel et al. [1-4], Moran et al. [5-7] and Borgnakke et al. [8-10], which are basically updated every 4 years. The latest textbook editions are the 9th edition by Çengel et al. in 2019 [4], the 9th edition by Moran et al. in 2018 [7] and the 10th edition by Borgnakke et al. in 2019 [10]. Textbooks in Chinese include [14, 15].

For the thermodynamic analyses of gas power cycles, the aforementioned textbooks mainly introduce the composition, performance parameters and performance analysis method for ideal Otto cycle, Diesel cycle, dual cycle and Brayton cycle. The change rule of performance parameters, such as thermal efficiency, with relevant parameters is presented. However, all textbooks, even monographs [16, 17], do not clearly point out: with certain conditions, such as the given cycle operating temperature limits, namely, the lower and upper limits of the cycle temperature (or inlet temperature and maximum cycle temperature), under what state should the gas power cycle work most economically? When this part of knowledge was learned by students in the class, these questions will be naturally raised: is the gas power cycle device with higher thermal efficiency the better economical one? What performance parameters should be used and aimed to design or evaluate gas power cycle devices in engineering applications?

The answer can not be found in textbooks or monographs, and will be presented in this work. As for a gas power cycle device within the given temperature limits, there exist a maximum specific cycle work and a maximum mean effective pressure for the reciprocating engines (basically piston-cylinder devices), corresponding to two different optimal compression ratios. All the aforementioned textbooks in English and in Chinese available do not systematically analyze the reciprocating gas power cycles, and only some textbooks involve in the analysis of Brayton cycle device. For example, the textbook [14] involves in the analysis of ideal gas turbine cycle. In the analysis of Brayton cycle device, the textbook [15] analyzes the conditions with the highest cycle thermal efficiency with an example, and gives the optimal compressor pressure ratio when the specific cycle work is the maximum. The textbooks [1-4] also involve in the analysis of ideal cycle of gas turbine, and give the optimal pressure ratio when the cycle specific power is maximum. The textbooks [5-7] also involve in the analysis of ideal cycle of gas turbine, and an example is used to derive the optimal pressure ratio when the cycle specific power is maximum. Wang made a detailed analysis in his monograph [18] on the most economical condition (involving only the maximum specific cycle work, not the maximum mean effective pressure) under which the gas power cycle device should be operated. However, the analyses of the optimal compression ratios are not presented by Wang [18] for the reciprocating devices with irreversible compression and expansion processes, and analyses for advanced efficient piston power cycles, such as Atkinson cycle and Miller cycles, are also not involved. This part should be introduced into the teaching process of Engineering Thermodynamics, and even be introduced into the textbooks of Engineering Thermodynamics. With this knowledge introduced, engineering applications can be better and clearer theoretically guided.

Instructors who are very familiar with the course of Engineering Thermodynamics are very clear: for any gas power cycle, as long as the infinitely increased pressure ratio, or compression ratio, namely the infinitely increased pressure or temperature at the end of the compression process, the theoretical maximum thermal efficiency can be infinitely close to the thermal efficiency of Carnot cycle operating within the same limits of temperature. Engine designed according to this guiding ideology has very high thermal efficiency, but the specific cycle work for an engine cycle is infinitesimal, close to zero. This is, obviously, not the goal to be pursued in engineering applications of gas power cycle devices, especially the moveable units. This implies that the traditional evaluation method of theory guidance for the engineering applications should be improved a lot.

In this paper, the authors concluded that it is necessary to present directly the performance parameters to evaluate correctly the gas power cycle devices in the textbook of Engineering Thermodynamics. That is to say, an ideal gas power cycle device should work at the condition of higher thermal efficiency, maximum specific cycle work, or maximum mean effective pressure, according to the applications. The actual gas power cycle device has optimal operating condition of the highest cycle thermal efficiency, the maximum specific cycle work, or the maximum mean effective pressure. The three conditions are not coincidental. The working condition depends on the nature (movable or stationary) of the device and its purpose.

Based on the generalized temperature-entropy diagram for gas power cycles, taking the ideal gasoline engine cycle, the Otto cycle, and the ideal pressurized Diesel cycle, the Miller-Diesel cycle, as examples, a new approach to the thermodynamic analysis with given limits of operation temperature is presented and illustrated for gas power cycles based on maximum specific cycle work, or maximum mean effective pressure. Textbook, organizing and planning the content with these presentations, can provide correct guidance to engineering applications.

**2. Generalized temperature-entropy diagram for gas power cycles**

According to a new framework system, i.e., Module, Project, Task and Subtask, a new textbook [19, 20] and the teaching manual [21] for the textbook have been edited in which the thermodynamic performance analyses of gas power cycles were newly presented. The optimal compression/pressure ratio, corresponding to the maximum specific cycle work or the maximum mean effective pressure were detailly analyzed and presented firstly for eight gas power cycles with the generalized *T-s* diagram, as shown in Fig. 1, including

1) **Otto cycle** which is the **Cycle 12341** with constant-volume heat addition (path 2-3) and constant-volume heat rejection (path 4-1);

2) **Atkinson cycle** which is the **Cycle 12341** with constant-volume heat addition (path 2-3) and constant-pressure heat rejection (path 4-1) is;

3) **Miller-Otto cycle** which is the **Cycle 123*b*41** with constant-volume heat addition (path 2-3), constant-volume heat rejection (path 4-*b*) and constant-pressure heat rejection (path *b*-1);

4) **Dual cycle** which is the **Cycle 12*a*341** with constant-volume heat addition (2-*a*), constant-pressure heat addition (path *a*-3) and constant-volume heat rejection (path 4-1);

5) **Diesel cycle** which is the **Cycle 12341** with constant-pressure heat addition (path 2-3) and constant-volume heat rejection (path 4-1);

6) **Miller-dual cycle** which is the **Cycle 12*a*34*b*1** with constant-volume heat addition (path 2-*a*), constant-pressure heat addition (path *a*-3), constant-volume heat rejection (path 4-*b*) and constant-pressure heat rejection (path *b*-1);

7) **Miller-Diesel cycle** which is the **Cycle 1234*b*1** with constant-pressure heat addition (path 2-3), constant-volume heat rejection (path 4-*b*) and constant-pressure heat rejection (path *b*-1) and

8) **Brayton cycle** which is the **Cycle 12341** with constant-pressure heat addition (path 2-3) and constant-pressure heat rejection (path 4-1).

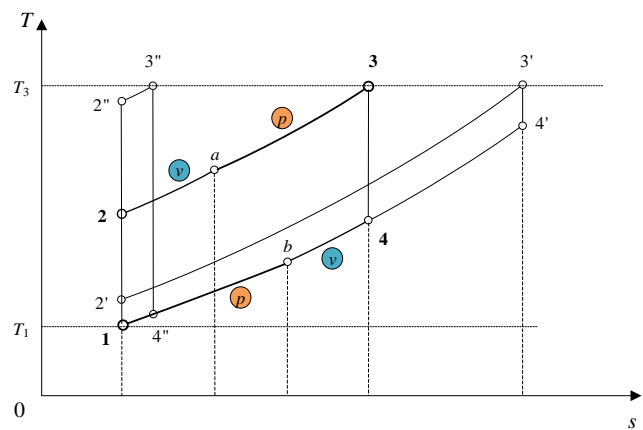

Fig. 1 Generalized *T-s* diagram for the gas power cycles [19-21].

The former 7 cycles are used for the reciprocating engines (basically piston-cylinder devices) and the last one, the Brayton cycle, is used for the gas-turbine engines. One thing must be clear that State 1, shown in Fig. 1, for each cycle is different depending on the cycles.

## 3. Otto cycle

The performance of reciprocating engines, such as Otto cycle, Diesel cycle, dual cycle, Atkinson cycle and Miller cycle engines, can be determined by applying the closed system energy balance and the second law along with property data including thermal efficiency, specific cycle work, mean effective pressure and the effects of varying compression ratio.

The cycle for spark-ignition reciprocating engines is idealized as the Otto cycle in the textbook of Engineering Thermodynamics. The cycle is shown in Fig. 2 (*p-v* diagram) and Fig. 3 (*T-s* diagram), including four reversible processes, i.e., path 1-2 (isentropic compression), path 2-3 (constant-volume heat addition), path 3-4 (isentropic expansion), and path 4-1 (constant-volume heat rejection).

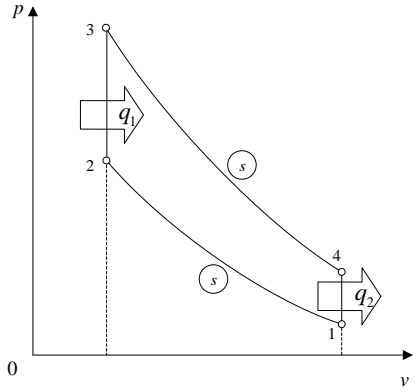 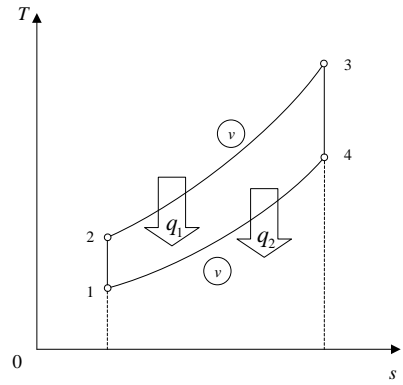

Fig. 2 *p-v* diagram for Otto cycle    Fig. 3 *T-s* diagram for Otto cycle

The cold-air-standard assumptions are used in the textbook of Engineering Thermodynamics and also used in this work. At this case, the properties of the working fluids [4] are, the gas constant $R_g$=0.2870 [kJkg-1K-1], the specific heat at constant pressure $c_p$=1.005 [kJkg-1K-1], the specific heat at constant volume $c_V$=0.718 [kJkg-1K-1] and the specific heat ratio $k$=1.400. When the compression ratio, $r = \frac{v_1}{v_2}$, and the maximum-to-minimum temperature ratio, $\tau = \frac{T_3}{T_1}$, are known, the Otto cycle is then determined. The expansion ratio, $r_E = \frac{v_4}{v_3}$, is same as the compression ratio for the Otto cycle. The thermal efficiency of an ideal Otto cycle is

$$\eta_{t,\text{Otto}} = 1 - \frac{1}{r^{k-1}} \qquad (1)$$

As can be seen from Eq. (1), the theoretical thermal efficiency of an ideal Otto cycle can be improved by increasing the compression ratio. At the condition of the determined maximum temperature in the cycle, the temperature of state 3 shown in Fig. 3, if the compression ratio is high enough (of course, it is restricted by various conditions in engineering, but it can be so imaged), the temperature of state 2, the end state of the compression, approaches to that of state 3, and that of state 4, the end state of the expansion, approaches to that of state 1. The Otto cycle will approach, or be very close to the Carnot cycle. The thermal efficiency of the Otto cycle engine is high enough, at this case, approaching to or being very close to the thermal efficiency of Carnot cycle working between the same temperature limit. Is this a perfect operation condition? Or should the future engineering applications work toward this condition? The current Engineering Thermodynamics textbooks do not theoretically answer this question positively, and students or engineers will naturally have such concerns when learning this part of knowledge.

Carefully analyzing the Otto cycle shown in Fig. 3, when the temperature limit of the cycle, i.e., the maximum cycle temperature $T_3$ and the minimum temperature $T_1$, is known, there exists a maximum specific cycle work corresponding to an optimal compression ratio. The *T-s* diagram of three Otto cycles with different compression ratios between $T_1$ and $T_3$ is shown in Fig. 4. The compression ratio of cycle 1→2'→3'→4'→1 is very small. It is not difficult to see that the thermal efficiency of this cycle is low, and the specific cycle work, i.e., the area enclosed by the process lines, is also very small. The compression ratio of cycle 1→2"→3"→4"→1 is quite large, with very high thermal efficiency (the limit is the efficiency of Carnot cycle in the same temperature limit) on the one hand, but the specific cycle work is very small (the limit is the net work done, or the specific cycle work, is zero) on the other hand. The compression ratio of cycle 1→2→3→4→1 is in the middle, and the thermal efficiency is in the middle, but the specific cycle work is relatively large. That is, with the change of compression ratio, there must exist a maximum specific cycle work between the lower and upper limits of absolute temperature, $T_1$ and $T_3$. Unfortunately, discussions about the maximum specific cycle work for an ideal Otto cycle between the determined values of $T_1$ and $T_3$ can not be found in any textbooks of Engineering Thermodynamics and the discussions will be presented in this work.

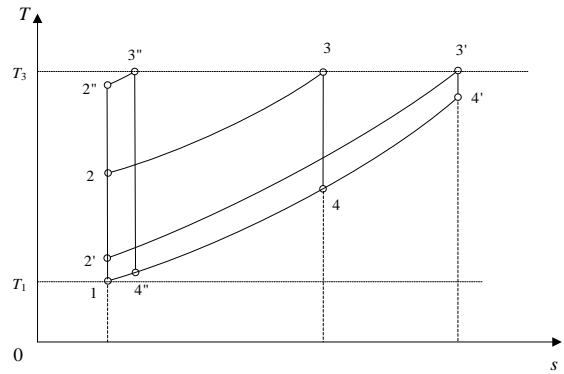

Fig. 4 *T-s* diagram for three Otto cycles with different compression ratios between $T_1$ and $T_3$

### 3.1 Optimal compression ratio

The specific cycle work for the Otto cycle with the cold-air-standard assumptions, shown in Figs. 2 and 3, can be expressed as

$$w_{net} = q_1 - q_2 = c_V(T_3 - T_2) - c_V(T_4 - T_1) = c_V(T_3 - T_2 - T_4 + T_1) \tag{2}$$

and the temperature of each state can be expressed as

$$T_2 = T_1\left(\frac{v_1}{v_2}\right)^{k-1} = T_1 r^{k-1}, \quad T_3 = \tau T_1, \quad T_4 = T_3\left(\frac{v_3}{v_4}\right)^{k-1} = \frac{T_1 T_3}{T_2}$$

Substituting these equations into the specific cycle work relation, Eq. (2), with the determined values of $T_1$ and $T_3$, and differentiating give

$$\frac{dw_{net}}{dT_2} = c_V\left(-1 + \frac{T_1 T_3}{T_2^2}\right) \tag{3}$$

$$\frac{d^2 w_{net}}{dT_2^2} = -2c_V \frac{T_1 T_3}{T_2^3} < 0 \tag{4}$$

Therefore, the specific cycle work does have a maximum value. Let Eq. (3) be zero, then the optimal end point temperature of compression is

$$T_{2,opt} = \sqrt{T_1 T_3} \tag{5}$$

Accordingly, when the specific cycle work is the maximum, the optimal end point temperature of expansion is

$$T_{4,opt} = \sqrt{T_1 T_3} = T_{2,opt} \tag{6}$$

An important conclusion is drawn based on Eq. (6): for an ideal Otto cycle, when the compression end point temperature, $T_2$, is exactly equal to the expansion end point temperature, $T_4$, the specific cycle work is the maximum value. In other words, when $T_2=T_4$, the maximum specific cycle work is reached for an Otto cycle. At this time, the optimal compression ratio is:

$$r_{w_{net}}^{opt} = \tau^{\frac{1}{2(k-1)}} \tag{7}$$

The reason that the maximum value of specific cycle work varies with the change of compression ratio is:

1) when the compression ratio is less than $r_{w_{net}}^{opt}$, the working fluid temperature of end point of expansion is always higher than that of the end point of compression, i.e., $T_4>T_2$. In this case, the increase of $q_1$ (the heat transfer to the working fluid) with the increment of $r$ is larger than that of $q_2$ (the heat transfer from the working fluid) with the increment of $r$, and the specific cycle work, $w_{net}$, increases with the increment of $r$.

2) when the compression ratio is larger than $r_{w_{net}}^{opt}$, the temperature of end point of expansion is always lower than that of the end point of compression, i.e., $T_4<T_2$. In this case, the increase of $q_1$ with the increment of $r$ is smaller than that of $q_2$, and the specific cycle work, $w_{net}$, decreases with the increment of $r$.

3) when the compression ratio is equal to $r_{w_{net}}^{opt}$, the temperature of end point of expansion is equal to that of the end point of compression, i.e., $T_4=T_2$. In this case, the increase of $q_1$ with the increment of $r$ is equal to that of $q_2$, and the specific cycle work, $w_{net}$, reaches its maximum.

### 3.2 Maximum specific cycle work

Substituting Eq. (5) for the temperature of the optimal end point of compression, and Eq. (6) for the temperature of the optimal end point of expansion equations into Eq. (2), the specific cycle work relation for the Otto cycle and simplifying give

$$w_{net,max} = c_V \left(\sqrt{T_3} - \sqrt{T_1}\right)^2$$

Or the dimensionless maximum specific cycle work

$$\frac{w_{net,max}}{c_V T_1} = \left(\sqrt{\frac{T_3}{T_1}} - 1\right)^2 = \left(\sqrt{\tau} - 1\right)^2 \tag{8}$$

This means that the dimensionless maximum specific cycle work is determined only by the temperature ratio, or, the maximum specific cycle work of an ideal Otto internal combustion engine is only a function of the specific heat at constant volume of the working fluid, the lower and upper limits of operating temperature, and is independent of the specific heat ratio. The maximum specific cycle work can be increased by raising the maximum operating temperature, lowering the minimum working fluid temperature and choosing the working fluid with larger specific heat at constant volume.

### 3.3 Maximum mean effective pressure

The mean effective pressure for the Otto cycle with the cold-air-standard assumptions, shown in Figs. 2 and 3, is defined as

$$p_{ME} = \frac{w_{net}}{v_1 - v_2} = \frac{w_{net}}{v_1\left(1 - \frac{1}{r}\right)} \tag{9}$$

With the former relations, Eq. (9) can be expressed as

$$p_{ME} = \frac{p_1}{k-1} \frac{\left(\frac{T_3}{T_2} - 1\right)\left(\frac{T_2}{T_1} - 1\right)}{1 - \left(\frac{T_2}{T_1}\right)^{\frac{1}{1-k}}} \tag{10}$$

or

$$p_{ME} = \frac{p_1}{k-1}\left[\frac{r(\tau+1)}{r-1} - \frac{\left(r^{2k-2} + \tau\right)}{(r-1)r^{k-2}}\right] \tag{11}$$

where, $k$ is the specific heat ratio. The mean effective pressure is found depending on the pressure of state 1, the specific heat ratio, the temperature ratio and the compression ratio. If $p_1$, $k$ and $\tau$ are fixed values, $p_{ME}$ depends only on the compression ratio and has extreme value.

Differentiating Eq. (11) with respect to $r$ and equating to zero give

$$(\tau+1) + (2k-2)r^{k-1}(r-1) = \left(r^{k-1} + \tau r^{1-k}\right)\left[(k-1)(r-1)+1\right] \tag{12}$$

This equation is implicit in $r$. Therefore, we have to use an equation solver or a trial-and-error approach to determine the compression ratio of the Otto cycle, $r_{p_{ME}}^{opt}$, at which the mean effective pressure reaches its maximum value, $p_{ME,max}$.

Furthermore, the power per liter, $P_L$(kWL-1), of a reciprocating engine at a specified revolution is related linearly with the mean effective pressure, as shown by Eq. (13).

$$P_L = \frac{n}{60000 n_{rev}} p_{ME} \tag{13}$$

where, $n$ is the revolutions per second of the engine and $n_{rev}$ is the revolutions for each thermodynamic cycle.

Eq. (13) indicates the compression ratio for the maximum power per liter is identical to that for maximum mean effective pressure.

**3.4 Optimal thermal efficiencies**

The optimal thermal efficiency corresponds to the optimal compression ratio at which the maximum specific cycle work or the maximum mean effective pressure is reached. Substituting the optimal compression ratio equation, Eq. (7), into the thermal efficiency relation, Eq. (1), and simplifying give the optimal thermal efficiency

$$\eta_{t,\text{Otto,opt}} = 1 - \frac{1}{\sqrt{\tau}} \tag{14}$$

This means the optimal thermal efficiency for the maximum specific cycle work is only the function of temperature ratio, $\tau=T_3/T_1$, independent of working fluid properties, and increases with the increase of $T_3/T_1$. That is, increasing $T_3$ or decreasing $T_1$ can improve the Otto engine thermal efficiency, which is consistent with the Carnot corollaries [5-7].

The analysis also shows that the specific cycle work of the Otto engine is inconsistent with its thermal efficiency, one of the traditional economic parameters. When the specific cycle work is the maximum, the thermal efficiency does not reach its maximum value. When the thermal efficiency approaches or reaches its maximum value, the specific cycle work approaches zero.

For the thermal efficiency when the maximum mean effective pressure is reached, can be obtained by substituting the solution of Eq. (12) into Eq. (1) and the analysis is omitted.

**3.5 Examples**

A simple ideal Otto cycle is taken as an example. The analysis shows the performance parameters of the cycle operating at a given compression ratio or an optimal compression ratio under given conditions, such as heat source temperature and environment temperature, especially the changes of specific cycle work, the thermal efficiency, the exergy destructions, and the second-law efficiency of this cycle.

**Example 1** An ideal air-standard Otto cycle has a compression ratio of 8. At the beginning of the compression process, air is at 100 [kPa] and 300 [K], and 800 [kJkg-1] of heat is transferred to air during the constant-volume heat-addition process from a source at 1900 [K] and waste heat is rejected to the surroundings at 300 [K]. Determine (a) the maximum temperature and pressure that occur during the cycle, (b) the specific cycle work and the thermal efficiency, (c) the mean effective pressure for the cycle, (d) the exergy destruction associated with each of the four processes and the cycle and (e) the second-law efficiency of this cycle.

**Assumptions 1** The air-standard assumptions are applicable. **2** Kinetic and potential energy changes are negligible. **3** Steady operating conditions exist.

**Analysis** The *p-v* and *T-s* diagrams of the ideal Otto cycle described is shown in Figs. 2 and 3. We note that the air contained in the cylinder forms a closed system.

(a) The maximum temperature and pressure in an Otto cycle occur at the end of the constant-volume heat-addition process (state 3). But first we need to determine the temperature and pressure of air at the end of the isentropic compression process (state 2) according to the air-standard assumptions and the process relations.

State 1

$$p_1 = 100, \quad T_1 = 300, \quad v_1 = \frac{R_g T_1}{p_1} = \frac{287.0 \times 300}{100 \times 10^3} = 0.8610$$

State 2 (isentropic compression of an ideal gas from state 1)

$$v_2 = \frac{v_1}{r} = \frac{0.8610}{8} = 0.1076, \quad T_2 = T_1 r^{k-1} = 300 \times 8^{1.4-1} = 689.22$$

$$p_2 = p_1 \left(\frac{v_1}{v_2}\right)^k = p_1 r^k = 100 \times 8^{1.4} = 1837.9$$

State 3 (constant-volume heat addition from state 2)

$$v_3 = v_2 = 0.1076, \quad q_1 = c_V (T_3 - T_2) = 0.718 \times (T_3 - 689.22) = 800$$

Thus

$$T_3 = 1803.42, \quad p_3 = p_2 \frac{T_3}{T_2} = 1837.9 \times \frac{1803.42}{689.22} = 4809.1$$

State 4 (isentropic expansion of an ideal gas from state 3)

$$v_4 = v_1 = 0.8610, \quad p_4 = p_3 \left(\frac{v_3}{v_4}\right)^k = 4.8091 \times \left(\frac{0.1076}{0.8610}\right)^{1.4} = 4809.1 \times \left(\frac{1}{8}\right)^{1.4} = 261.7$$

$$T_4 = \frac{p_4 v_4}{R_g} = \frac{261.7 \times 10^3 \times 0.8610}{287.0} = 784.99$$

(b) the specific cycle work and the thermal efficiency

The heat transfer from the working fluid is

$$q_2 = c_V (T_4 - T_1) = 0.718 \times (784.99 - 300) = 348.22$$

The specific cycle work for one cycle is

$$w_{net} = q_1 - q_2 = 800 - 348.22 = 451.78$$

The thermal efficiency of the cycle is determined from its definition

$$\eta_{th} = \frac{w_{net}}{q_1} = \frac{451.78}{800} = 0.5647$$

Under the cold-air-standard assumptions, the thermal efficiency would be

$$\eta_{t,Otto} = 1 - \frac{1}{r^{k-1}} = 1 - \frac{1}{8^{1.4-1}} = 0.5647$$

(c) The mean effective pressure is determined from its definition, Eq. (9)

$$p_{ME} = \frac{w_{net}}{v_1 - v_2} = \frac{451.78}{0.8610 - 0.1076} = 599.7$$

(d) The exergy destructions

Processes 1-2 and 3-4 are isentropic ($s_1=s_2$, $s_3=s_4$) and therefore do not involve any internal or external irreversibilities; that is, $x_{dest,12} = 0$ and $x_{dest,34} = 0$.

Processes 2-3 and 4-1 are constant-volume heat-addition and heat-rejection processes, respectively, and are internally reversible. However, the heat transfer between the working fluid and the source or the sink takes place through a finite temperature difference, rendering both processes irreversible. The exergy destruction associated with each process is determined. However, first we need to determine the entropy change of air during these processes:

The entropy change for process 2-3 of ideal gas

$$s_3 - s_2 = c_V \ln \frac{T_3}{T_2} = 0.718 \times \ln \frac{1803.42}{689.22} = 0.6906$$

Also
$$q_1=1000 \text{ and } T_H=1900$$
Thus
$$x_{dest,23} = T_0\left[(s_3 - s_2)_{sys} - \frac{q_1}{T_H}\right] = 300\text{ K} \times \left(0.6906 - \frac{800}{1900}\right) = 80.87$$

The average temperature of air in the constant-volume heat-addition is
$$\bar{T}_1 = \frac{q_1}{s_3 - s_2} = \frac{800}{0.6906} = 1158.36$$

For process 4-1, $s_4 - s_1 = s_2 - s_3 = -0.6906$, $q_{41} = q_2 = 348.22$, and $T_{sink} = 300$. Thus,
$$x_{dest,\,41} = T_0\left[(s_1 - s_4)_{sys} + \frac{q_{41}}{T_{sink}}\right] = 300 \times \left[-0.6906 + \frac{348.22}{300}\right] = 141.03$$

Therefore, the irreversibility of the cycle is
$$x_{dest,cycle} = x_{dest,12} + x_{dest,23} + x_{dest,34} + x_{dest,41} = 0 + 80.87 + 0 + 141.03 = 221.90$$

Notice that the largest exergy destruction in the cycle occurs during the heat-rejection process. Therefore, any attempt to reduce the exergy destruction, or improve the performance, should start with this process.

The average temperature of air in the constant-volume heat-rejection is
$$\bar{T}_2 = \frac{q_2}{s_3 - s_2} = \frac{348.22}{0.6906} = 504.21$$

The thermal efficiency of the cycle is determined from the average temperature
$$\eta_{t,Otto} = 1 - \frac{\bar{T}_2}{\bar{T}_1} = 1 - \frac{504.21\text{ K}}{1158.36\text{ K}} = 0.5647$$

(e) The second-law efficiency is defined as
$$\eta_{II} = \frac{\text{Exergy recovered}}{\text{Exergy expended}} = \frac{x_{recovered}}{x_{expended}} = 1 - \frac{x_{destroyed}}{x_{expended}}$$

Here the expended energy is the energy content of the heat supplied to the air in the engine (which is its work potential) and the energy recovered is the net work output
$$x_{expended} = x_{heat,in} = \left(1 - \frac{T_0}{T_H}\right)q_{in} = \left(1 - \frac{300}{1900}\right) \times 800 = 673.68, \quad x_{recovered} = w_{net} = 451.78$$

Substituting, the second-law efficiency of this cycle is determined to be
$$\eta_{II} = \frac{x_{recovered}}{x_{expended}} = \frac{451.78}{673.68} = 0.6706 \text{ or } \eta_{II} = 1 - \frac{x_{destroyed}}{x_{expended}} = 1 - \frac{221.90}{673.68} = 0.6706$$

**Example 2** If the ideal air-standard Otto cycle operates between the lowest temperature $T_1=300$ [K], and $T_3=1803.42$ [K], and at the compression ratio of optimal instead of 8, repeat Example 1.

**Assumptions 1** The air-standard assumptions are applicable. **2** Kinetic and potential energy changes are negligible. **3** Steady operating conditions exist.

**Analysis** The *p-v* and *T-s* diagrams of the ideal Otto cycle described is shown in Figs. 2 and 3. We note that the air contained in the cylinder forms a closed system. The optimal compression ratio should be firstly determined with the given temperature limit.

From Eq. (7), the optimal compression ratio is

$$r_{w_{net}}^{opt} = \left(\frac{T_3}{T_1}\right)^{\frac{1}{2(k-1)}} = \left(\frac{1803.42}{300}\right)^{\frac{1}{2\times(1.4-1)}} = 9.4128$$

(a) The maximum temperature and pressure in an Otto cycle occur at the end of the constant-volume heat-addition process (state 3). The temperature is known at state 3. The pressure of air at the end of the isentropic compression process (state 2) can be determined as

State 1
$$p_1 = 100, \quad T_1 = 300, \quad v_1 = 0.8610$$

State 2(isentropic compression of an ideal gas from state 1)
$$v_2 = \frac{v_1}{r_{w_{net}}^{opt}} = \frac{0.8610}{9.4128} = 0.09147, \quad T_2 = T_1\left(r_{w_{net}}^{opt}\right)^{k-1} = 300 \times 9.1428^{1.4-1} = 735.55$$

$$p_2 = p_1\left(\frac{v_1}{v_2}\right)^k = p_1\left(r_{w_{net}}^{opt}\right)^k = 100 \times 9.4128^{1.4} = 2307.8$$

State 3(constant-volume heat addition from state 2)
$$v_3 = v_2 = 0.09147, \quad q_1 = c_V(T_3 - T_2) = 0.718 \times (1803.42 - 735.54) = 766.74$$

$$p_3 = p_2 \frac{T_3}{T_2} = 2307.8 \times \frac{1803.42}{735.54} = 5658.4$$

State 4(isentropic expansion of an ideal gas from state 3)
$$v_4 = v_1 = 0.8610, \quad p_4 = p_3\left(\frac{v_3}{v_4}\right)^k = 5658.4 \times \left(\frac{1}{9.1248}\right)^{1.4} = 245.2$$

$$T_4 = \frac{p_4 v_4}{R_g} = \frac{245.2 \times 10^3 \times 0.8610}{287.0} = 735.55 = T_2$$

(b) the specific cycle work and the thermal efficiency

The heat transfer from the working fluid is
$$q_2 = c_V(T_4 - T_1) = 0.718 \times (735.55 - 300) = 312.72$$

The specific cycle work for one cycle is
$$w_{net} = q_1 - q_2 = 766.74 - 312.72 = 454.02$$

The thermal efficiency of the cycle is determined from its definition
$$\eta_{th} = \frac{w_{net}}{q_1} = \frac{454.02}{766.74} = 0.5921$$

Under the cold-air-standard assumptions, the thermal efficiency would be
$$\eta_{t,Otto} = 1 - \frac{1}{\left(r_{w_{net}}^{opt}\right)^{k-1}} = 1 - \frac{1}{9.4128^{1.4-1}} = 0.5921$$

(c) The mean effective pressure is determined from its definition
$$p_{ME} = \frac{w_{net}}{v_1 - v_2} = \frac{454.02}{0.8610 - 0.09147} = 590$$

The compression ratio at which the maximum mean effective pressure is reached is determined from Eq. (12). By trial-and-error approach, we have
$$r_{p_{ME}}^{opt} = 5.2179$$

And from Eq. (11), we then have $p_{ME,max}$ =609, which is larger than the former values of the mean effective pressure. At this case, the thermal efficiency will be decreased since the compression ratio is lowered.

(d) The exergy destructions

The exergy destructions can be determined following the ways shown in example 1.

The entropy change for process 2-3 of ideal gas is

$$s_3 - s_2 = c_V \ln \frac{T_3}{T_2} = 0.718 \times \ln \frac{1803.42}{735.55} = 0.6439$$

Also

$$q_1 = 766.74 \text{ and } T_H = 1900$$

Thus

$$x_{dest,23} = T_0 \left[ (s_3 - s_2)_{sys} - \frac{q_1}{T_H} \right] = 300 \times \left( 0.6439 - \frac{766.74}{1900} \right) = 72.11$$

The average temperature of air in the constant-volume heat-addition is

$$\bar{T}_1 = \frac{q_1}{s_3 - s_2} = \frac{766.74}{0.6439} = 1190.78$$

For process 4-1, $s_4 - s_1 = s_2 - s_3 = -0.6439$, $q_{41} = q_2 = 312.72$, and $T_{sink} = 300$. Thus,

$$x_{dest,41} = T_0 \left[ (s_1 - s_4)_{sys} + \frac{q_{41}}{T_{sink}} \right] = 300 \times \left[ -0.6439 + \frac{312.72}{300} \right] = 119.55$$

Therefore, the irreversibility of the cycle is

$$x_{dest,cycle} = x_{dest,12} + x_{dest,23} + x_{dest,34} + x_{dest,41} = 0 + 72.11 + 0 + 119.55 = 191.66$$

Also, notice that the largest exergy destruction in the cycle occurs during the heat-rejection process. Therefore, any attempt to reduce the exergy destruction, or improve the performance, should start with this process.

The average temperature of air in the constant-volume heat-rejection is

$$\bar{T}_2 = \frac{q_2}{s_3 - s_2} = \frac{312.72}{0.6439} = 485.67$$

The thermal efficiency of the cycle is determined from the average temperature

$$\eta_{t,Otto} = 1 - \frac{\bar{T}_2}{\bar{T}_1} = 1 - \frac{485.67}{1190.78} = 0.5921$$

(e) The second-law efficiency is

$$x_{expended} = \left( 1 - \frac{T_0}{T_H} \right) q_{in} = \left( 1 - \frac{300}{1900} \right) \times 766.74 = 645.68$$

$$x_{recovered} = w_{net} = 454.02$$

Substituting, the second-law efficiency of this cycle is determined to be

$$\eta_{II} = \frac{x_{recovered}}{x_{expended}} = \frac{454.02}{645.68} = 0.7032 \quad \text{or} \quad \eta_{II} = 1 - \frac{x_{destroyed}}{x_{expended}} = 1 - \frac{191.66}{645.68} = 0.7032$$

Comparing the results from Examples 1 and 2, the Otto engine operates between the lower limit of temperature $T_1 = 300$ [K] and the upper limit of working temperature $T_3 = 1803.42$ [K] having two determined optimal compression ratios and calculated by Eq. (7) and Eq. (12), respectively. When the Otto cycle operates at the compression ratio of 8, as shown in example 1, the specific cycle work, the thermal efficiency and the second-law efficiency are 451.78 [kJkg-1], 56.47% and 67.06%, respectively. While, when it operates at the optimal compression ratio of 9.4128, as shown in example 2, the specific cycle work, the thermal efficiency and the second-law efficiency are 454.02 [kJkg-1], 59.21% and 70.32%, respectively. Obviously, the performance of the Otto engine operates at the optimal compression ratio determined by the lower and upper limits of absolute temperature is better than that operates at compression ratio of 8 in the same temperature limit. The Otto cycle operates at the optimal compression ratio does the maximum specific cycle work. The thermal efficiency and the second-law efficiency are not always increased, although in the examples they are. When the optimal compression ratio is larger than the original compression ratio, the efficiencies at the optimal compression ratio will increase, otherwise, they will decrease.

Working at the optimal compression ratio, even though the efficiency of an ideal Otto cycle is decreased (when $r_{w_{\text{net}}}^{\text{opt}} > r$ it will increase), the specific cycle work is maximized, and then the equipment can be miniaturized; or, with the device of the same size, the net output work will be larger. That is, the Otto cycle operating at the optimal compression ratio is more economical than that operating at other compression ratios. In other words, when an Otto cycle operating at $r < r_{w_{\text{net}}}^{\text{opt}}$ is switched to the optimal compression ratio, the overall economy of the device is better as a result of the improvement of optimal compression ratio of the device, resulting the increased thermal efficiency, the specific cycle work and the exergy efficiency. On the whole, it is more economical and reasonable for the Otto cycle to operate at the optimal compression ratio.

## 4. Ideal Miller-Diesel cycle

Generally, the efficiency and the net work output for an engine can be improved by lowering the compression ratio and/or increasing the expansion ratio. According to this idea, Miller [22] proposed the use of early intake valve closing to provide internal cooling before compression so as to reduce compression work. A modified Diesel cycle is known as the Miller-Diesel cycle whose *p-v* and *T-s* diagrams are shown in Figs. 5 and 6. Compared to the Diesel cycle, the Miller-Diesel cycle has a shorter effective compression stroke (path 1-2) and the same expansion stroke (path 3-4). By analogy, a modified Otto cycle is known as the Miller-Otto cycle and a modified dual cycle is known as the Miller-dual cycle.

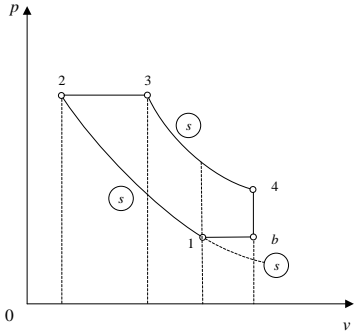  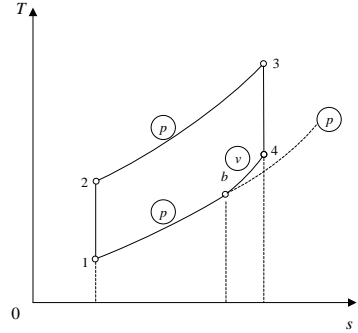

Fig. 5 *p-v* diagram for the ideal Miller-Diesel cycle    Fig. 6 *T-s* diagram for the ideal Miller-Diesel cycle

By the way, the analyses for Miller cycles which have been widely used in engineering are rarely found in the textbooks.

For the determined lower and upper limits of temperature, the ideal Miller-Diesel cycle is determined by the compression ratio $r=v_1/v_2$ (smaller than the Diesel cycle) and the expansion ratio $r_E=v_4/v_3$ (the same as the Diesel cycle). When an ideal Miller-Diesel cycle is analyzed on a cold air-standard basis, the specific heats are taken as constant. Kinetic and potential energy effects are negligible.

The temperature of each state can be expressed as

$$T_2 = T_1\left(\frac{v_1}{v_2}\right)^{k-1} = T_1 r^{k-1}, \quad T_3 = \tau T_1, \quad T_4 = T_3\left(\frac{v_3}{v_4}\right)^{k-1} = \frac{T_3}{r_E^{k-1}},$$

$$T_b = T_1 \frac{v_b}{v_1} = T_1 \frac{v_4}{v_1} = T_1 \frac{v_4}{v_3}\frac{v_3}{v_2}\frac{v_2}{v_1} = \frac{T_1 T_3}{T_2}\frac{r_E}{r} = r_E T_3 \frac{T_1^{\frac{k}{k-1}}}{T_2^{\frac{k}{k-1}}}$$

The amount of heat added to the cycle is

$$q_1 = c_p (T_3 - T_2) \tag{15}$$

The total amount of heat rejected from the cycle is

$$q_2 = c_V (T_4 - T_b) + c_p (T_b - T_1) \tag{16}$$

Then, the thermal efficiency is determined from its definition to be

$$\eta_{t,\text{Miller-Diesel}} = 1 - \frac{\dfrac{\tau}{kr_E^{k-1}} + \dfrac{k-1}{k}\dfrac{r_E \tau}{r} - 1}{\tau - r^{k-1}} \tag{17}$$

Like the Otto cycle shown in Figs. 3 and 4, when the lower and upper limits of temperature and the expansion ratio of an ideal Miller-Diesel cycle are known, there exists a compression ratio at which the specific cycle work reaches maximum, or a compression ratio at which the mean effective pressure reaches maximum. This idea should be known by students in the Engineering Thermodynamics class.

**4.1 Optimal compression ratio**

The specific cycle work for an ideal Miller-Diesel cycle with the cold-air-standard assumptions, shown in Figs. 5 and 6, can be expressed as

$$\begin{aligned} w_{\text{net}} &= q_1 - q_2 = c_p(T_3 - T_2) - c_V(T_4 - T_b) - c_p(T_b - T_1) \\ &= c_V \left[ k(T_3 - T_2) - (k-1)T_b - T_4 + kT_1 \right] \end{aligned} \tag{18}$$

Substituting equations of the state temperatures into the specific cycle work relation, Eq. (13), yields

$$w_{\text{net}} = c_V \left[ k(T_3 - T_2) - (k-1) r_E T_3 T_1^{\frac{k}{k-1}} T_2^{-\frac{k}{k-1}} - \frac{T_3}{r_E^{k-1}} + kT_1 \right] \tag{19}$$

With the determined limits of temperature and the specified expansion ratio, the specific cycle work depends only on the end temperature of compression, $T_2$. Differentiating with respect to $T_2$ gives

$$\frac{dw_{\text{net}}}{dT_2} = c_V \left[ -k + k r_E T_3 T_1^{\frac{k}{k-1}} T_2^{-\frac{k}{k-1}-1} \right] \tag{20}$$

$$\frac{d^2 w_{\text{net}}}{dT_2^2} = c_V k r_E T_3 T_1^{\frac{k}{k-1}} \left( -\frac{2k-1}{k-1} \right) T_2^{-\frac{k}{k-1}-2} < 0 \tag{21}$$

This implies the specific cycle work reaches its maximum value at the end temperature of compression

$$T_{2,w_{\text{net}}}^{\text{opt}} = \left(r_E T_3 T_1^{\frac{k}{k-1}}\right)^{\frac{k-1}{2k-1}} = T_b \tag{22}$$

This indicates that the specific cycle work for a Miller-Diesel cycle will be maximum when the end temperature of the compression is equal to the end temperature of the isochoric heat rejection. Notice that Eq. (22) for the ideal Miller-Diesel cycle is different from that of the ideal Otto cycle as shown by Eq. (6). The compression ratio corresponding to maximum specific cycle work is the optimal pressure ratio.
Thus

$$r_{w_{\text{net}}}^{\text{opt}} = \left(r_E \tau\right)^{\frac{1}{2k-1}} \tag{23}$$

**4.2 Maximum specific cycle work**

Substituting Eq. (22) for the end temperature of the optimal compression into Eq. (19) and simplifying give the dimensionless maximum specific cycle work for the ideal Miller-Diesel

$$\frac{w_{\text{net,max}}}{c_V T_1} = k(\tau+1) - 2\sqrt{k(k-1)(r_E \tau)^{\frac{2(k-1)}{2k-1}}} - \frac{\tau}{r_E^{k-1}} \tag{24}$$

This indicates the maximum specific cycle work of the Miller-Diesel cycle is only a function of the specific heat at constant volume, the specific heat ratio, the expansion ratio and the temperature ratio of the cycle. The maximum cyclic specific work can be increased by increasing the maximum working temperature and/or the expansion ratio, decreasing the minimum working medium temperature, and selecting a working fluid with a large specific heat at constant volume or a high specific heat ratio.

**4.3 Maximum mean effective pressure**

The mean effective pressure for the Miller-Diesel cycle with the cold-air-standard assumptions, shown in Figs. 5 and 6, is defined as

$$p_{\text{ME}} = \frac{w_{\text{net}}}{v_1 - v_2} = \frac{w_{\text{net}}}{v_1\left(1 - \frac{1}{r}\right)} \tag{25}$$

With the former relations, Eq. (25) can be expressed as

$$p_{\text{ME}} = \frac{p_1}{k-1} \frac{k\left(\tau - \frac{T_2}{T_1}\right) - (k-1)r_E \tau T_1^{\frac{k}{k-1}} T_2^{-\frac{k}{k-1}} - \frac{\tau}{r_E^{k-1}} + k}{1 - \left(\frac{T_2}{T_1}\right)^{\frac{1}{1-k}}} \tag{26}$$

or

$$p_{\text{ME}} = \frac{p_1}{k-1}\left[\frac{k\left(\tau - r^{k-1}\right) - (k-1)r_E \tau \frac{1}{r^k} - \frac{\tau}{r_E^{k-1}} + k}{1 - \frac{1}{r}}\right] \tag{27}$$

The mean effective pressure is found depending on the pressure of state 1, the specific heat ratio,

the temperature ratio, the compression ratio and the expansion ratio. If $p_1$, $k$, $r_E$ and $\tau$ are fixed values, $p_{ME}$ depends only on the compression ratio and has extreme value.

Differentiating Eq. (27) with respect to $r$ and equating to zero give

$$k^2 r^{k-1} + r^k \left(1-k^2\right) - (k-1)^2 r_E \tau r^{-k}(r-1) + (k-1) r_E \tau r^{1-k} = k(\tau+1) - \frac{\tau}{r_E^{k-1}} \tag{28}$$

This equation is implicit in $r$. Therefore, we have to use an equation solver or a trial-and-error approach to determine the compression ratio of the Miller-Diesel cycle, $r_{p_{ME}}^{opt}$, at which the mean effective pressure reaches its maximum value, $p_{ME,max}$.

Furthermore, the power per liter, $P_L$(kWL-1), of a reciprocating engine at a specified revolution is related linearly with the mean effective pressure, as shown by Eq. (29).

$$P_L = \frac{n}{1000 n_{rev}} p_{ME} \tag{29}$$

This implies that the compression ratio for the maximum power per liter is identical to that for maximum mean effective pressure.

## 4.4 Optimal thermal efficiencies

The optimal thermal efficiency corresponds to the optimal compression ratio at which the maximum specific cycle work or the maximum mean effective pressure is reached. Substituting the optimal compression ratio equation, Eq. (23), into the thermal efficiency relation, Eq. (17), and simplifying give the optimal thermal efficiency

$$\eta_{t,\text{Miller-Diesel,opt}} = 1 - \frac{\frac{\tau}{k r_E^{k-1}} + \frac{k-1}{k}(r_E \tau)^{\frac{2(k-1)}{2k-1}} - 1}{\left[\tau - (r_E \tau)^{\frac{k-1}{2k-1}}\right]} \tag{30}$$

This means the optimal thermal efficiency for the maximum specific cycle work is the function of temperature ratio, the specific heat ratio and the expansion ratio, and increases with the increase of $T_3/T_1$. That is, increasing $T_3$ or decreasing $T_1$ can improve the Miller-Diesel engine thermal efficiency, which is consistent with the Carnot corollaries [5-7].

The analysis also shows that the specific cycle work of the Miller-Diesel engine is inconsistent with its thermal efficiency, one of the traditional economic parameters. When the specific cycle work is the maximum, the thermal efficiency does not reach its maximum value. When the thermal efficiency approaches or reaches to its maximum value, the specific cycle work approaches zero.

For the optimal thermal efficiency for the maximum mean effective pressure can be obtained by substituting the solution of Eq. (28) into Eq. (17) and the analysis is omitted.

Therefore, there should be a compromise between the compression ratio, thus the thermal efficiency, the net work output and the mean effective pressure (or the maximum power per liter) according to the application areas of reciprocating engines. With less work output per cycle, a larger mass flow rate, thus a larger system, is needed to maintain the same power output, which may be uneconomic.

## 5. Conclusion remarks

Gas power cycles in the course of Engineering thermodynamics are the important content of practical application of basic theory and principle of Engineering Thermodynamics. The textbooks introduce mainly the composition, the performance parameters (such as the thermal efficiency) and the performance analysis method for the ideal Otto, the Diesel, the dual, as well as the Brayton cycles. The variation of thermal efficiency with relevant parameters is analyzed and discussed.

However, the maximum value of specific cycle work or mean effective pressure and its boundary conditions for the gas power cycles, especially the reciprocating engines, are rarely discussed and no recommendations for different usages (movable or fixation) of gas power cycle devices are presented for their economic operation parameters in the textbooks of the currently edited Engineering Thermodynamics. Therefore, engineering applications of gas power cycle devices cannot be well directed. After learning this part of course contents of the currently edited Engineering Thermodynamics, students/engineers are difficult to abandon the inveteracy idea that the more efficient power cycle device is the better gas power cycle device.

Taking devices of ideal Otto cycle or ideal Miller-Diesel cycle as examples, this paper analyzes the optimum (economical) compression ratios at which the specific cycle net power, or the mean effective pressure, is maximum. Following the same idea presented in this work, the optimal compression/pressure ratio for the operation condition of maximum specific cycle work, or maximum mean effective pressure, for other gas power cycles can also be derived at the known conditions of operation temperature limits and other parameters necessary to form the cycle, including ideal and actual cycles. For the analyses of actual cycles, only the irreversibilities of compression and expansion processes are considered. There exists also the operating condition of maximum efficiency for an actual cycle, besides that of the maximum specific cycle work or the maximum mean effective pressure, and these three conditions are not the same. According to the using purpose of the device, the design and transformation of the device should be based on the maximum specific cycle work/mean effective pressure (such as movable equipment) or the maximum cycle thermal efficiency (such as fixation equipment), rather than taking the thermal efficiency as the only performance parameter. In engineering applications, this should be the theoretical guidelines for the design of gas power cycle devices.

For evaluations of the gas power cycle devices of movable utilization, two very important concepts, i.e., the maximum specific cycle work and the maximum mean effective pressure, and then the optimal compression/pressure ratios, should be emphasized. The maximum specific cycle work/mean effective pressure, instead of the cycle thermal efficiency, should be pursued. However, the higher cycle thermal efficiency should be pursued for gas power cycle devices of fixation utilization. Designing or modifying or retrofitting gas power devices this way, the total economy of the device can be reached and the theoretical guidance of Engineering Thermodynamics to engineering practice can be truly reflected. Therefore, the concepts of maximum specific cycle work or mean effective pressure, and then the optimal compression/pressure ratios (or optimal thermal efficiencies) are suggested to be introduced to the teaching process of Engineering Thermodynamics, finally introduced to the currently edited textbooks when they are revised and reprinted in the future.

This paper presents a new approach for the thermodynamic analyses of gas power cycles. The authors suggest the textbooks of Engineering Thermodynamics should pay more attentions to the two important concepts of the maximum specific cycle work and the maximum mean effective pressure, and then the optimal compression/pressure ratios, or the optimal thermal efficiencies at the maximum values, in addition to its thermal efficiency, so as to realize clear and correct guidance of the theory and principles of Engineering Thermodynamics to engineering applications and better services to engineering practices.

**Nomenclature**

| | | | |
|---|---|---|---|
| $c_p$ [Jkg-1K-1] | Specific heat at constant pressure | $r_E$ [-] | Expansion ratio |
| $c_V$ [Jkg-1K-1] | Specific heat at constant volume | $s$ [kJkg-1K-1] | Specific entropy |
| $k$ [-] | Specific heat ratio | $T$ [K] | Temperature |
| $\dot{m}$ [kgs-1] | Mass flow rate | $v$ [m3kg-1] | Specific volume |
| $n$ [revs-1] | Revolutions per second | $w_{net}$ [kJkg-1] | Specific net work for a cycle |
| $n_{rev}$ [rev cycle-1] | Revolutions per cycle | $x$ [kJkg-1] | Specific exergy |
| $p$ [kPa] | Pressure | | |
| $P_L$ [kWL-1] | Power per liter | Greek symbols | |
| $p_{ME}$ [kPa] | Mean effective pressure for a cycle | $\tau$ [-] | The maximum-to-minimum temperature ratio for a cycle |
| $q$ [kJkg-1] | Specific heat addition or heat rejection | $\pi$ [-] | Pressure ratio |
| $R_g$ [kJkg-1K-1] | Gas constant | $\eta$ [-] | efficiency |
| $r$ [-] | Compression ratio | | |

**Ethical statement**

   Not Applicable

**Consent statement**

   Not Applicable